\documentclass[twocolumn,prl,floatfix,superscriptaddress,footinbib]{revtex4-1}
\usepackage{hyperref}
\usepackage{amsmath}
\usepackage{amssymb}
\usepackage{graphicx}
\usepackage{bm}
\usepackage{color}

\newcommand{\editP}[1] {\textcolor{black}{#1}}
\newcommand{\editR}[1] {\textcolor{black}{#1}}

\begin{document}

\title{Non-Hermitian topology: a unifying framework for the Andreev versus Majorana states controversy}

\author{J. Avila}
\thanks{These authors contributed equally to this work}
\affiliation{Materials Science Factory, Instituto de Ciencia de Materiales de Madrid (ICMM), Consejo Superior de Investigaciones Cient\'{i}ficas (CSIC), Sor Juana In\'{e}s de la Cruz 3, 28049 Madrid, Spain}
\author{F. Pe\~naranda}
\thanks{These authors contributed equally to this work}
\affiliation{Departamento de F\'{i}sica de la Materia Condensada, Condensed Matter Physics Center (IFIMAC) and Instituto Nicol\'{a}s Cabrera, Universidad Aut\'{o}noma de Madrid, E-28049 Madrid, Spain}
\author{E.~Prada}
\affiliation{Departamento de F\'{i}sica de la Materia Condensada, Condensed Matter Physics Center (IFIMAC) and Instituto Nicol\'{a}s Cabrera, Universidad Aut\'{o}noma de Madrid, E-28049 Madrid, Spain}
\author{P.~San-Jose}
\email{pablo.sanjose@csic.es}
\affiliation{Materials Science Factory, Instituto de Ciencia de Materiales de Madrid (ICMM), Consejo Superior de Investigaciones Cient\'{i}ficas (CSIC), Sor Juana In\'{e}s de la Cruz 3, 28049 Madrid, Spain}
\author{R.~Aguado}
\email{raguado@icmm.csic.es}
\affiliation{Materials Science Factory, Instituto de Ciencia de Materiales de Madrid (ICMM), Consejo Superior de Investigaciones Cient\'{i}ficas (CSIC), Sor Juana In\'{e}s de la Cruz 3, 28049 Madrid, Spain}

\date{\today}
\begin{abstract}
Andreev bound states (ABSs) in hybrid semiconductor-superconductor nanowires can have near-zero energy in parameter regions where band topology predicts trivial phases. This surprising fact has been used to challenge the interpretation of a number of transport experiments in terms of non-trivial topology with Majorana zero modes (MZMs). 
We show that this ongoing ABS versus MZM controversy is fully clarified when framed in the language of non-Hermitian topology, the natural description for open quantum systems. This change of paradigm allows us to understand topological transitions and the emergence of pairs of zero modes more broadly, in terms of exceptional point (EP) bifurcations of system eigenvalue pairs in the complex plane. Within this framework, we show that some zero energy ABSs are actually non-trivial, and share all the properties of conventional MZMs, such as the recently observed $2e^2/h$ conductance quantization. From this point of view, any distinction between such ABS zero modes and conventional MZMs becomes artificial. The key feature that underlies their common non-trivial properties is an asymmetric coupling of Majorana components to the reservoir, which triggers the EP bifurcation. 
\end{abstract}\maketitle
\emph{Introduction}---Since the remarkable prediction \cite{Lutchyn:PRL10,Oreg:PRL10} that a hybrid semiconductor-superconductor nanowire can be tuned into a topological superconductor phase with MZMs \cite{Kitaev:PU01}, there have been a number of papers reporting experimental data in the form of a zero-bias anomaly (ZBA) in the differential conductance ($dI/dV$) for increasing Zeeman fields \cite{Mourik:S12,Deng:S16,Zhang:NC17,Suominen:PRL17,Nichele:PRL17,Aguado:RNC17, Lutchyn:NRM18}. This behavior is consistent with tunneling into a MZM that emerges after the system undergoes a topological phase transition. 

This Majorana interpretation has recently been challenged since an alternative explanation in terms of ABSs with near-zero energy in the topological trivial phase, namely for Zeeman fields smaller than the critical field predicted by band topology $B<B_c$, reproduces all the expected phenomenology in transport. Following early calculations that proved that smooth confinement potentials inevitably lead to near-zero energy ABSs \cite{Kells:PRB12,Prada:PRB12}, a number of papers  \cite{Liu:PRB17,Moore:PRB18,Setiawan:PRB17,Moore:18,Liu:18,Vuik:18} have reported numerical observations that systematically demonstrate that, indeed, ABSs in the trivial regime mimic Majoranas. This nagging ABS-versus-MZM question is compounded by the recent observation of $2e^2/h$ conductance quantization \cite{Zhang:N18}, which can be also reproduced by ABSs  \cite{Moore:18}, and thus is considered a serious objection in the field. 
\begin{figure}
\centering \includegraphics[width=\columnwidth]{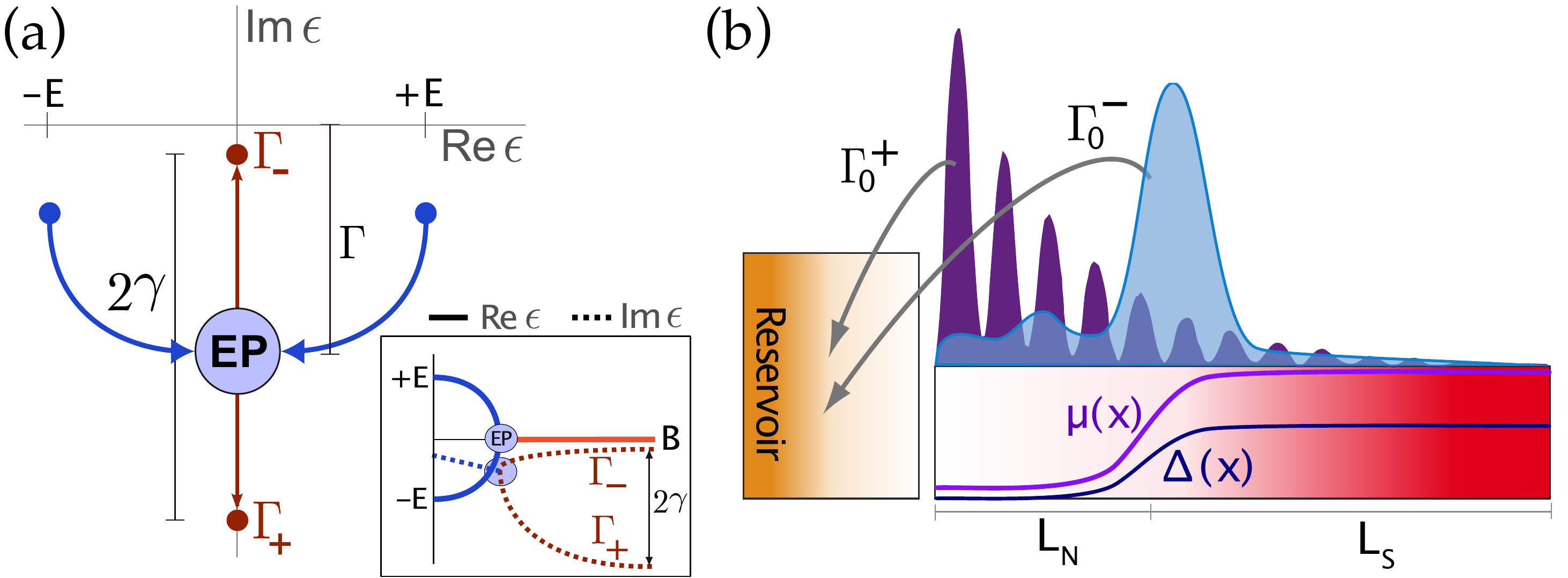}
\caption{\label{Fig1} \textbf{Exceptional points.} (a) The bifurcation, as a function of some external parameter $B$, of two complex Green's function poles across an exceptional point (EP) estabilises quasi-bound Majorana zero modes (red). Inset shows the evolution of real and imaginary pole energies across the EP. (b) Sketch of the normal-superconductor (NS) junction formed when a proximitized nanowire with inhomogeneous chemical potential and pairing, $\mu(x)$ and $\Delta(x)$, is coupled to a reservoir. Such junction is a natural host for Majorana zero modes even below the critical field $B<B_c$ that emerge from EP bifurcations around parity crossings when their coupling is asymmetric $\Gamma_0^+>\Gamma_0^-$ due to spatial non-locality.}
\end{figure}

In this letter, we argue that the above question is ill posed, and is the result of a viewpoint, that of band topology, only truly applicable to semi-infinite systems. We argue that a more general framework, relevant to the experimental setup and rigorously well defined for finite samples, allows us to precisely distinguish trivial from non-trivial zero modes. Among the latter are the MZMs from conventional band topology theory, but also a large subset of ABSs zero modes. From this point of view both kinds of states are really one and the same, which explains why they cannot be distinguished. The key idea to understand this claim is to realize that, instead of conventional band topology, the natural language to describe a normal-superconductor (NS) junction, the geometry relevant to transport experiments, is that of open quantum systems. In particular, we consider the non-Hermitian topology defined in terms of the complex poles $\epsilon_p$ of the retarded Green's function (or, equivalently, of the scattering matrix)
\begin{equation}
G^r(\omega)=[\omega-H_\mathrm{eff}(\omega)]^{-1}.
\end{equation}
Here $H_\mathrm{eff}(\omega)=H_0+\Sigma(\omega)$ is an effective non-Hermitian Hamiltonian which takes into account \emph{both} the system (through $H_0$) and its coupling to the reservoir (through the self-energy $\Sigma(\omega)$). The poles of the retarded Green's function can be viewed as complex eigenvalues of $H_\mathrm{eff}(\omega)$ and have a well defined physical interpretation that generalizes the spectrum of the isolated system (namely, real eigenvalues of $H_0$). They define quasi-bound states in the open system, with complex energies $\epsilon_p=E-i\Gamma$, that decay to the reservoir with rate $\Gamma\geq0$. As discussed first by Pikulin and Nazarov \cite{Pikulin:JL12, Pikulin:PRB13}, 
the distribution of complex eigenvalues of $H_\mathrm{eff}$ allows for a natural topological classification of open system phases. This generalises that of band topology, defined solely in terms of $H_0$. 

\emph{Exceptional points and non-Hermitian topology}---
In a closed, semi-infinite, quasi-1D superconducting system, with a bulk described by the Bloch Hamiltonian $H_0(\mathbf{k})$, a non-trivial topological invariant for $H_0$ rigorously implies that a protected Majorana zero mode should arise at the system boundary, by virtue of the bulk-boundary correspondence. This conventional band-topological picture cannot be invoked in finite systems, or to establish the protection or lack thereof of ABSs zero modes when the bulk $H_0$ is trivial. In finite-length systems the problem of discerning between ABSs zero modes and MZMs is actually ambiguous, as the wavefunctions of both states are continuously connected, and topological transitions take the form of mere crossovers. The protection of zero modes is no longer an all-or-nothing proposition in finite systems, but a matter of degree, ultimately connected to the degree of wavefunction non-locality of the zero mode in question \cite{Prada:PRB17,Deng:A17}.

The change to an open setting with a non-Hermitian $H_\mathrm{eff}$ has deep implications, and in particular allows us to recover a precise topological criterion to distinguish trivial from non-trivial zero modes by including the reservoir into the problem. When coupled to a metallic reservoir, a zero mode or parity crossing of the closed system may or may not become stabilised at zero energy, transforming into a robust zero bias anomaly in transport, insensitive to perturbations. Stabilisation of this kind provides the precise criterion for topologically non-trivial zero modes.
%
The correct language to understand the stabilisation mechanism is that of bifurcations of the complex eigenvalues. 
These are a direct consequence of the underlying charge-conjugation symmetry of the Bogoliubov-de Gennes formalism, which dictates that if $\epsilon$ is an eigenvalue, so is $-\epsilon^*$. This condition 
can be satisfied in two non-equivalent ways, see Fig.\ref{Fig1}a. One can have \emph{pairs} of eigenvalues 
located symmetrically at opposite sides of the imaginary axis (blue dots) 
or, alternatively, have \emph{independent} self-conjugate eigenvalues 
lying exactly on the imaginary axis (red dots). The former correspond to standard finite-energy ABSs (Bogoliubov excitations symmetrically located at $\pm E$ and with equal decay rate $\Gamma$ to the reservoir). The latter correspond to non-trivial zero modes in the context of open systems. A bifurcation of two trivial ABSs ($\epsilon_\pm = -\epsilon_\mp^*=\pm E - i\Gamma$) into two non-trivial zero modes with different decay rates ($\epsilon_\pm =-\epsilon_\pm^*= - i\Gamma_\pm$) defines an exceptional point (EP).

More generally, EPs are points in parameter space where a non-Hermitian $H_\mathrm{eff}$ becomes non-diagonalizable, and are well studied in the non-Hermitian quantum physics literature. EPs have been extensively discussed in the context of open photonic systems, see e.g. \cite{Zhen:N15}, and, more recently, in the context of non-Hermitian topology  \cite{Leykam:PRL17,Shen:PRL18,Kozii:17,Gong:18}, which is the relevant scenario here. In our context, EPs generalise and extend the concept of topological transitions of conventional band topology. Crucially for the issue at hand, while a non-trivial band topology in the closed system does imply the emergence of non-trivial open-system zero modes, the converse is not true: a trivial system in the band-topological sense can develop EPs and protected zero modes when it is opened to a reservoir. Most of the so-called ``trivial'' $B<B_c$ ABS zero modes in the literature are of this kind, and are thus just as non-trivial as MZMs within this language.

Let us illustrate the mathematical structure of an EP bifurcation by considering the low energy Hamiltonian of a single parity crossing, $H_0=E_0\tau_z$, with $\tau_z$ the Pauli matrix in the $(c^\dagger, c)$ particle-hole space. Since, mathematically, one can always decompose a local ABS quasiparticle excitation in terms of two Majorana operators, $\gamma_{\pm} \sim c \pm c^\dagger$, it is enlightening to write the previous Hamiltonian in the Majorana basis and take into account the possibility, allowed by charge-conjugation symmetry, that each of these Majoranas is coupled \emph{differently} to the reservoir, with couplings $\Gamma_0^-\neq\Gamma_0^+$. The Hamiltonian in the Majorana basis reads:
\begin{equation}
 \label{EP}
H_M= \left( \begin{array}{ccc}
-i\Gamma_0^- & -i E_0\\
i E_0& -i\Gamma_0^+  \end{array} \right),
\end{equation}
Its eigenvalues are $\epsilon_\pm= -i\Gamma_0\pm\sqrt{E_0^2-\gamma_0^2} = E_\pm - i\Gamma_\pm$, in terms of the average \emph{coupling} $\Gamma_0\equiv (\Gamma_0^+ + \Gamma_0^-)/2$ and its asymmetry $\gamma_0\equiv (\Gamma_0^+-\Gamma_0^-)/2$. The square root term produce two different regimes. For $|E_0|>\gamma_0$ we obtain the standard ABS solution with opposite real energies $E_\pm = \pm\sqrt{E_0^2-\gamma_0^2}$ and equal decays $\Gamma_\pm = \Gamma_0$. In contrast, when 
$|E_0|<\gamma_0$ we get two purely imaginary eigenvalues, $E_\pm=0$, with different decay to the reservoir, $\Gamma_\pm= \Gamma_0\pm \sqrt{\gamma_0^2-E_0^2}$. The two regimes are separated by the EP bifurcation where the square root vanishes, and are characterised by zero/non-zero normalised \emph{decay} asymmetry $\gamma/\Gamma = (\Gamma_+ - \Gamma_-)/(\Gamma_+ + \Gamma_-)$
\footnote{The generality of the above effective model goes beyond Majorana physics and can be encountered in many other settings, such as e.g. two-orbital Dirac materials with two distinct quasiparticle lifetimes \cite{Kozii:17}, where an explicit connection between EPs and Fermi arcs can be made, disordered Weyl semimetals \cite{Zyuzin:18}, dissipative cold atomic gases \cite{Xu:18}, or in PT-symmetric photonic systems \cite{Eleuch:AP14,Zhang:SA18}.}. 

\begin{figure} 
\centering \includegraphics[width=\columnwidth]{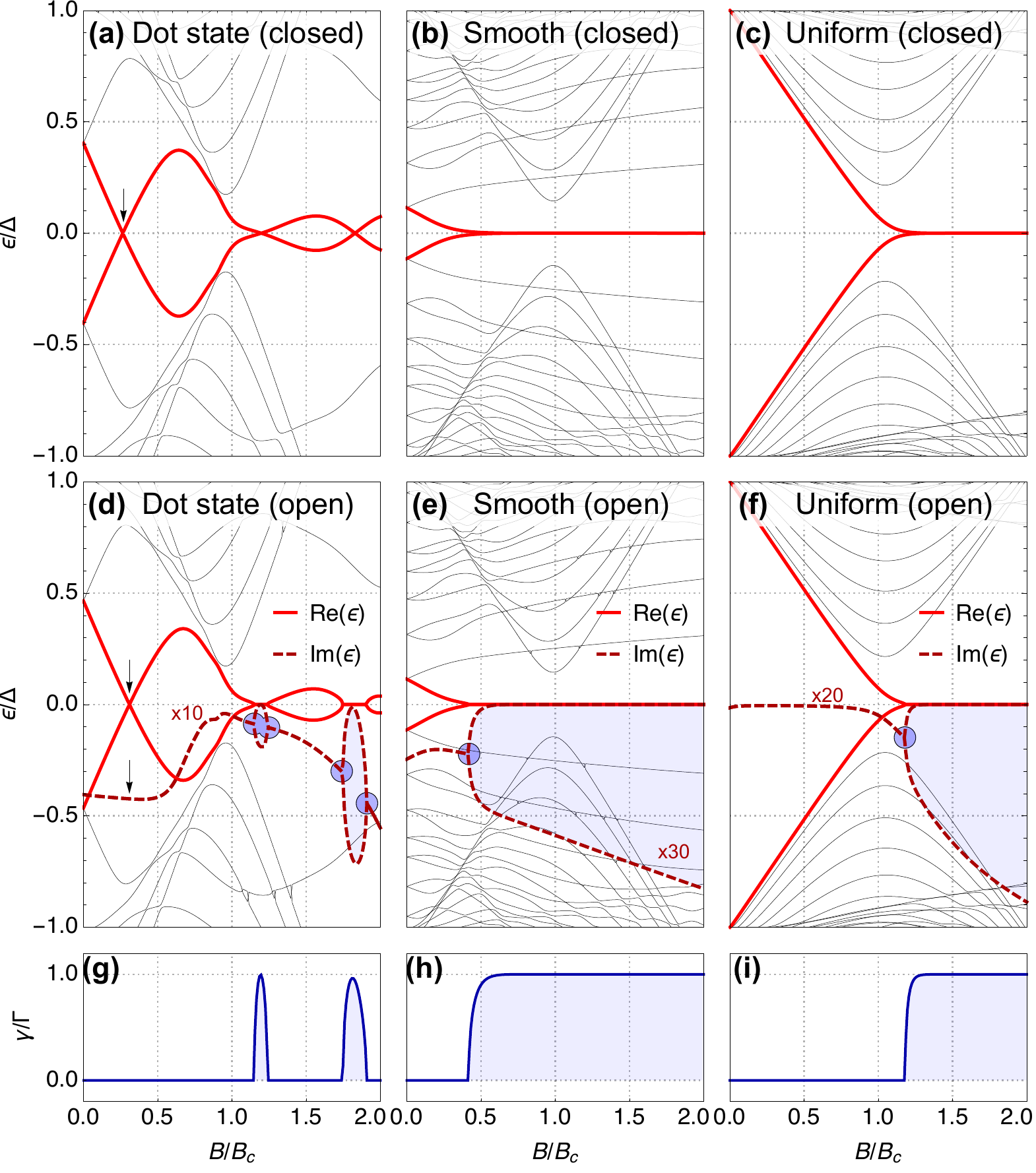}
\caption{\label{Fig2} \textbf{Trivial and non-trivial zero modes}. The spectrum versus Zeeman field of isolated Rashba nanowires [with a quantum dot state in a short wire (a), with smoothly confined ABS (b), and a uniform case (c)] becomes complex when coupled to a metallic reservoir (d,e,f). The decay rates (imaginary energy) of the two lowest states are shown in dashed red. Exceptional points (circles) appear as decay rate bifurcation accompanied by real energies stabilised at zero, and may arise for $B<B_c$ [panel (e)], or more conventionally for $B\geq B_c$ [panels (d,f)]. Panels (g,h,i) show the decay asymmetry $\gamma/\Gamma$ that defines non-Hermitian non-triviality. See detailed parameters in the Supplementary Information (Table I).}
\end{figure}

\emph{Trivial and non-trivial ABS zero modes}--- We now discuss two archetypical instances of trivial and non-trivial ABS zero modes in open systems: a quantum dot parity crossing \cite{Lee:NN14} and a smoothly confined ABS. Both exist in the trivial $B<B_c$ band-topological phase of the Lutchyn-Oreg microscopic model for a Rashba nanowire (see Refs. \cite{Lutchyn:PRL10, Oreg:PRL10} for details) in the presence of spatially inhomogeneous potentials $\Delta(x)$, $\mu(x)$, that define a normal region on the left end, see Fig. \ref{Fig1}b,
\begin{figure*}
\centering \includegraphics[width=\textwidth]{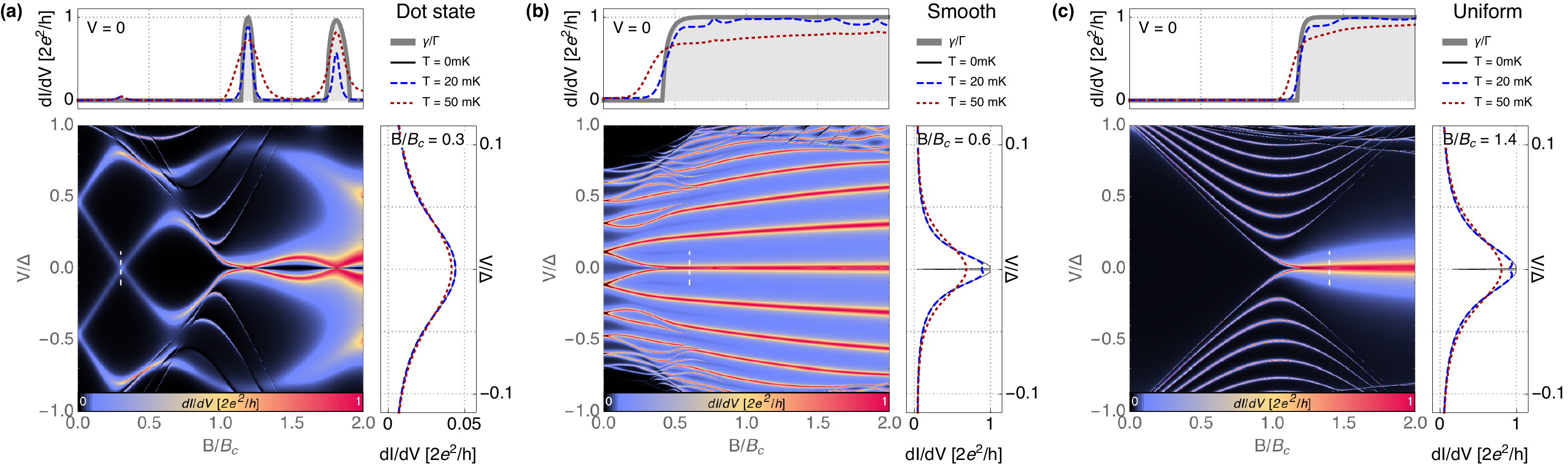}
\caption{\label{Fig3} \textbf{Differential conductance across an EP}. $dI/dV$ as a function of bias $V$ and Zeeman field $B$ for the three systems of Fig. \ref{Fig2}: short wire with a quantum dot state, nanowire with a smoothly confined ABS and a nanowire with uniform density and pairing. Top panels show the $dI/dV$ at $V=0$ at low temperatures ($T=20$mK and $T=50$mK, blue and red dashed lines, respectively). These jump towards a quantised $2e^2/h$ value just as the Majorana asymmetry (thick grey line) becomes finite $\gamma/\Gamma\sim 1$ upon crossing an EP {(non-trivial topology)}. Right panels show the ZBAs at fixed $B$ (see white dashed cuts in the density plots). Only the latter two (non-trivial $\gamma/\Gamma>0$) reach $2e^2/h$ at $T\to 0$.}
\end{figure*}
\begin{equation}
H_0 = \left(\frac{p_x^2}{2m^*} - \mu(x)\right) \tau_z + B\sigma_x\tau_z + \frac{\alpha}{\hbar} p_x \sigma_y \tau_z + \Delta(x)\tau_x.
\label{Lutchyn-Oreg}
\end{equation}
For the quantum dot case, the normal region is taken much shorter than the coherence length $\xi\sim\hbar v_F/\Delta$ and is weakly connected to the nanowire. {It hosts a quantum dot-like state with \emph{spatially local} Majorana components, and hence a symmetric coupling to the reservoir $\gamma_0/\Gamma_0\approx 0$. For the smoothly confined case, the normal region is comparable or larger than $\xi$, and is connected to the nanowire by a smoothly varying $\mu(x)$ and $\Delta(x)$. It hosts near-zero ABS with substantially \emph{non-local} Majorana components, and hence $\gamma_0/\Gamma_0>0$ (see wavefunctions in Fig \ref{Fig1}b and Supp. Inf. for further analysis)}. The corresponding spectrum of the two isolated systems {(decoupled from the reservoir)} as a function of Zeeman field $B$ is shown in Figs. \ref{Fig2}(a,b), respectively, with the conventional case of a long and uniform nanowire shown in panel (c) for comparison. In the former cases, we note that a zero mode appears for $B<B_c$, either in the form of a parity crossing (quantum dot case, black arrow) \cite{Lee:NN14} or as a robust zero energy mode (smooth case). Upon coupling the left end of the nanowire to a featureless metallic reservoir,
the $H_\mathrm{eff}(B)$ eigenvalues become complex. The real and imaginary parts of the lowest-lying levels are shown in panels (d-f) as solid and dashed red lines, respectively. In panel (d) we see that the fully local quantum dot state, with zero coupling asymmetry $\gamma_0/\Gamma_0=0$, is not stabilised for $B<B_c$, and remains as a point-like parity crossing in the real spectrum, with a single finite lifetime. It is therefore a \emph{trivial} ABS. Conversely, and despite their apparent similarity, the Majorana parity crossings at $B>B_c$ {that result from the short nanowire length} are \emph{non trivial}, with bifurcating lifetimes (blue circles). The smoothly confined ABS, panel (e), similarly shows a bifurcation of its two Majorana decay rates $\Gamma_\pm$, and becomes stabilised at zero real energy. It is thus revealed as a non-trivial zero mode appearing \emph{well before} the critical Zeeman field $B_c$. After the EP, the two Majorana quasi-bound zero modes are indistinguishable from the conventional $B>B_c$ MZMs of the uniform nanowire, panel (f).  
The physically relevant property of these zero modes is their decay asymmetry $\gamma/\Gamma$, shown in panels (g-i). As in the effective model, phases with non-trivial (trivial) non-Hermitian topology are defined by $\gamma/\Gamma>0$ ($\gamma/\Gamma=0$). 

An important phenomenon usually takes place after an EP whereby the decay rate of one of the zero modes becomes vanishingly small ($\Gamma_-\to 0$). 
Since the decay asymmetry dictates a constraint on the timescales of key non-trivial properties like non-Abelian braiding or the $4\pi$-periodic Josephson effect, such pole decoupling has practical importance, as it enables e.g. adiabatic braiding operations. 
Pole decoupling, however, is merely a crossover, and is distinct in this framework from the actual topological transition at the well-defined EP. 

\emph{Conductance quantization after an EP bifurcation}--- We now show how EPs are directly observable in transport. In Fig. \ref{Fig3} we present the typical behaviour of the differential conductance $dI/dV$ for the dot {coupled to a short nanowire} (a), the smooth case (b), and the uniform nanowire (c), corresponding to the {systems} in Figs. \ref{Fig2}(d,e,f). In the top panels we see that, as soon as the system crosses an EP and the decay asymmetry jumps to a non-trivial 
 $\gamma/\Gamma\sim 1$ (thick gray lines), the low-temperature linear conductance $dI/dV|_{V\to 0}$ becomes nearly quantised to $2e^2/h$  (results for $T=20$mK and $T=50$mK are shown as blue and red dashed lines, respectively). 
\editR{At zero temperature and constant $B$ (white dashed cuts in the density plots), these $2e^2/h$ transport anomalies show, as a function of bias $V$, a characteristic split-Lorentzian profile [right panels in Figs. \ref{Fig3}(b,c)], indistinguishable in the smooth and uniform cases. This structure is a signature of the bifurcated poles, and hence of non-trivial topology, with the widths of the broader peak and central dip corresponding to $\Gamma_+$ and $\Gamma_-$, respectively. It results from Fano interference of transport through the two zero modes that emerge from the EP. The $2e^2/h$ peak height is a manifestation of perfect Andreev reflection into the strongly coupled zero mode (width $\Gamma_+$) at small bias $|V|>\Gamma_-$. Charge-conjugation symmetry dictates that the tunneling amplitudes for electrons and holes are equal, thus reflecting the particle-equals-antiparticle principle of the bifurcated Majorana zero modes.}


This well-known result \cite{Eleuch:AP14,Ioselevich:NJP13} has recently been rediscovered \cite{Moore:18,Vuik:18} but the important connection with the EP physics discussed here has thus far been overlooked.
This connection naturally explains why zero modes at $B<B_c$ systematically result in $2e^2/h$-quantised ZBAs expected at $B>B_c$, as soon as temperature exceeds $\Gamma_-$. After the pole decoupling $\Gamma_-\to 0$ there is no way to distinguish between a smoothly confined $B<B_c$ ABS, a finite-length $B>B_c$ MZM, or a MZM in a strictly semi-infinite $B>B_c$ nanowire. They all exhibit a low temperature differential conductance of $2e^2/h$, independently of any fine tuning. In particular it cannot exceed $2e^2/h$, in contrast to the $4e^2/h$ of standard ABSs in the limit of perfect Andreev reflection. 

To conclude, we have shown that by adopting the language of non-Hermitian topology of open systems, the topological nature of zero energy states in superconducting nanowires is clarified. The complex energy of zero modes with an asymmetric coupling to the reservoir bifurcate at EPs, which define an open-system generalisation of the band-topological transitions of close systems. Thus, so called trivial ABSs may bifurcate into quasi-bound zero modes, exactly like conventional MZMs, as long as their coupling asymmetry to the reservoir  is larger than their energy $\gamma_0>|E_0|$. This crucial asymmetry requirement is automatically satisfied 
as soon as the wavefunctions of the zero-mode's Majorana components become spatially separated, as we demonstrate in the Supp. Inf. Such separation, known as Majorana non-locality, spontaneously develops in smoothly confined ABS zero modes, which hence become topological zero modes in this context, indistinguishable from MZMs in all their properties, including their $2e^2/h$ quantized conductance. 
Experimentally, we expect smoothly confined ABSs to be a common occurrence in clean samples, which explains the ubiquity of robust zero bias anomalies for $B<B_c$ {(see smoothness analysis in the Supp. Inf.)}.
We also speculate that
braiding schemes based on the ability to couple and manipulate individual Majoranas (through e.g. measurement-based braiding \cite{Bonderson:PRB13,Plugge:NJP17,Karzig:PRB17}) should also be possible using the non-Hermitian MZMs discussed here.  
While a finite Majorana non-locality is the universal and experimentally relevant mechanism to achieve EP-mediated topological protection of zero modes in an open setting, it is important to stress that reservoir engineering could also be used to stabilise zero modes that are originally local in the closed system. This has been explicitly demonstrated for trivial zero-energy parity crossings that become stable zero modes when coupled to a spin-polarised reservoir \footnote{In this latter case, even fully local Majoranas may exhibit a $4\pi$-Josephson effect and parametric non-Abelian braiding \cite{San-Jose:SR16}.}.  Unlike EPs arising from spatial non-locality, however, such spin-selective schemes do not guarantee that the stabilised zero modes enjoy generic protection against decoherence. 

\vspace{12pt}
\small{
\noindent\textbf{Acknowledgements}\\
We thank J. Cayao for useful discussions in the early stages of this work. Research supported by the Spanish Ministry of Economy and Competitiveness through Grants FIS2015-65706-P, FIS2015-64654-P, FIS2016-80434-P (AEI/FEDER, EU), the Ram\'on y Cajal programme Grants RYC-2011-09345, RYC-2013-14645, and the Mar\'ia de Maeztu Programme for Units of Excellence in R\&D (MDM-2014-0377). 

\section{Supplemental Information}

\subsection{Methods} The results presented in the main text (Figs. 2 and 3) were obtained using the MathQ software~\cite{MathQ} to simulate a nanowire described by the discretized Lutchyn-Oreg model in Eq.\ref{Lutchyn-Oreg}, where $\sigma_{x,y,z}$ and $\tau_{x,y,z}$ are Pauli matrices for the spin and electron-hole degrees of freedom, $m^*$ is the semiconducting quasiparticle effective mass, $\alpha$ is the spin-orbit coupling and $B$ is the induced Zeeman field. We consider position-dependent chemical potentials and induced pairing,
\begin{eqnarray}
\mu(x)&=&\mu_N+(\mu_B-\mu_N)\,\theta_{\zeta_B}(x-L_B)\nonumber\\
&&+(\mu_S-\mu_B)\,\theta_{\zeta_N}(x-L_B-L_N),\nonumber\\
\Delta(x)&=&\Delta\,\theta_{\zeta_S}(x-L_B-L_N).
\end{eqnarray}
Here the smoothed-out step functions are defined as $\theta_\zeta(x) = \frac{1}{2}[1+\tanh(x/\zeta)]$, where $\zeta$ is the step width. $L_N$, $L_B$, and $L_S$ correspond to the lengths of the normal, barrier and proximitised regions, respectively (see  Fig. \ref{Fig1 SI} (a) for the quantum dot case and (b) for the smooth one). The nanowire is coupled on the left ($x=0$) to a featureless reservoir (constant density of states) with chemical potential $\mu_{res}\gg\mu_N,\mu_S$ (in all the calculations, we fix this chemical potential to be $\mu_{res}$=5meV). The tuneable coupling at $x=0$ is modeled by an adimensional parameter $\tau\in[0,1]$ which renormalizes the original tight-binding hopping parameter $t=\hbar^2/2m^*a_0^2$, with $m^*$=0.015$m_e$ (corresponding to InSb nanowires),
$m_e$ the electron's mass and $a_0$ a lattice discretization parameter. The coupling of $H_0$ to the reservoir creates a self-energy at $x=0$, which defines our effective non-Hermitian Hamiltonian $H_\mathrm{eff}$.
Materials parameters of the physical systems discussed in the main text are $\Delta$=0.5 meV, $\alpha=0.5\,\mathrm{eV\AA}$ (which gives a spin-orbit length $l_{SO}=\hbar^2/(m^*\alpha)\sim$102nm), while the different geometries are constructed with parameters given in Table \ref{tab:params}.

\begin{figure} 
\centering \includegraphics[width=\columnwidth]{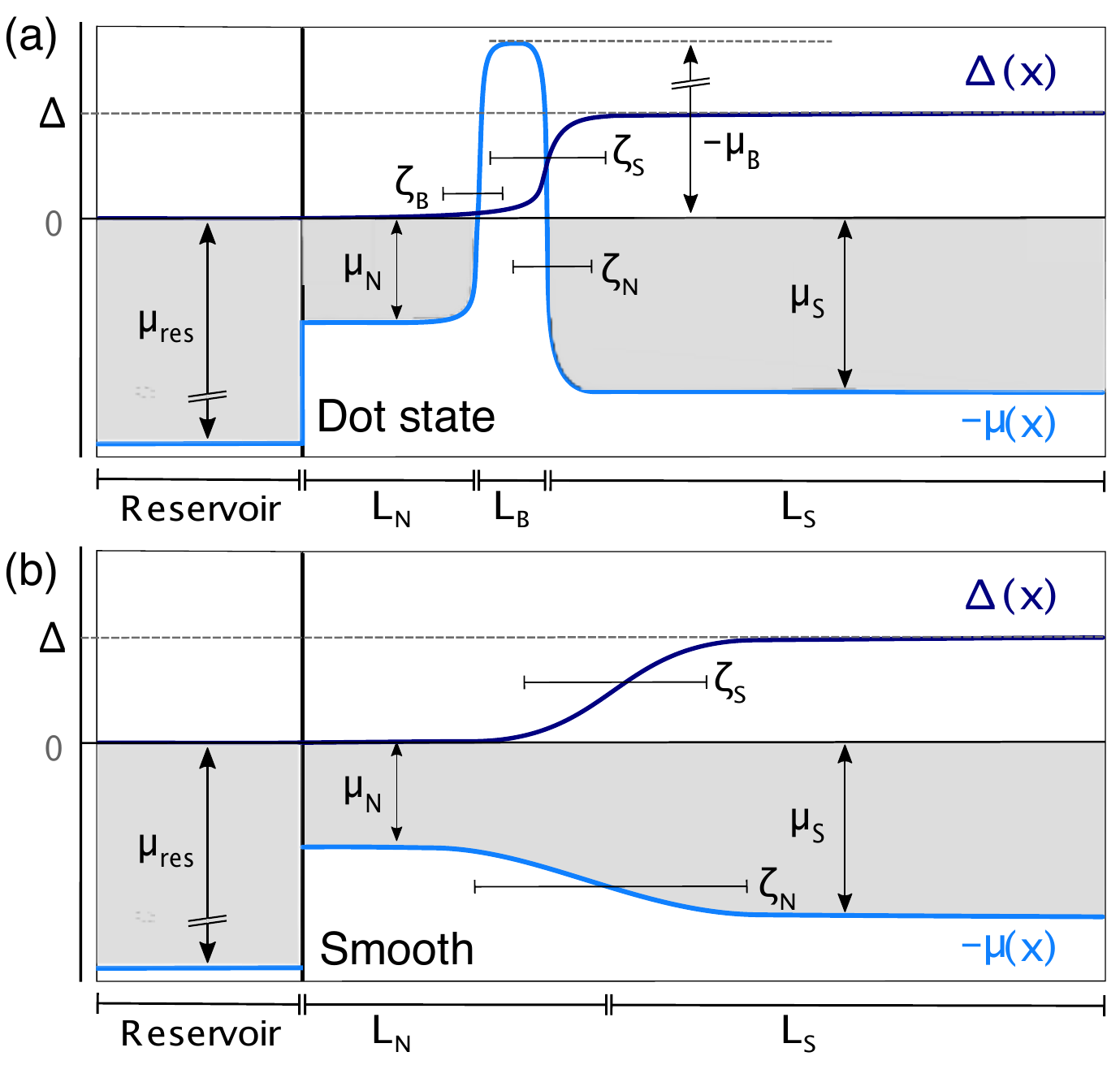}
\caption{\label{Fig1 SI} \textbf{Schematics of a smooth NS junction and relevant length scales}. The position dependent Fermi energy $\mu(x)$ and pairing $\Delta(x)$ define four regions in our general nanowire model: normal reservoir, normal region of length $L_N$, barrier of length $L_B$ and proximitised region $L_S$, with smooth transitions of length $\zeta_{B,N,S}$ between each. The reservoir contact transparency is tuneable. The three systems studied here (dot, smooth, uniform) differ only in their parameters, see Table \ref{tab:params}. Panel (a) shows the quantum dot case and panel (b) the smooth junction case.}
\end{figure}

\begin{table}[h!]
\caption{\textbf{Model parameters} for the three nanowire models studied in this work.}
\begin{center}
\begin{tabular}{ |c|c|c|c|}
  \hline
  parameter & quantum dot & smooth & uniform\\ 
    \hline
  $\mu_N$[meV]& $0.9$ & $0.1$ & 0.0\\
  \hline
   $\mu_S$[meV] & $0.88$& $0.3$ & 0.0\\
  \hline
  $\mu_B$[meV] & $-9.0$& 0.0& 0.0\\
  \hline
  $\Delta$ [meV]& 0.5 & 0.5 & 0.5\\
  \hline
  $\alpha$ [eV\AA]& 0.5 & 0.5 & 0.5\\
  \hline
   $L_N[\mu$m] & 0.15 & 1.0& 0.0\\
  \hline
   $L_S[\mu$m] & 1.0 & 2.0 &  2.0\\
  \hline
  $L_B[\mu$m]& 0.02 & 0.0 & 0.0\\
  \hline
  $\zeta[\mu$m]& 0.0 & $0.3$& 0.0\\
  \hline
\end{tabular}
\end{center}
\label{tab:params}
\end{table}
\begin{figure} 
\centering \includegraphics[width=\columnwidth]{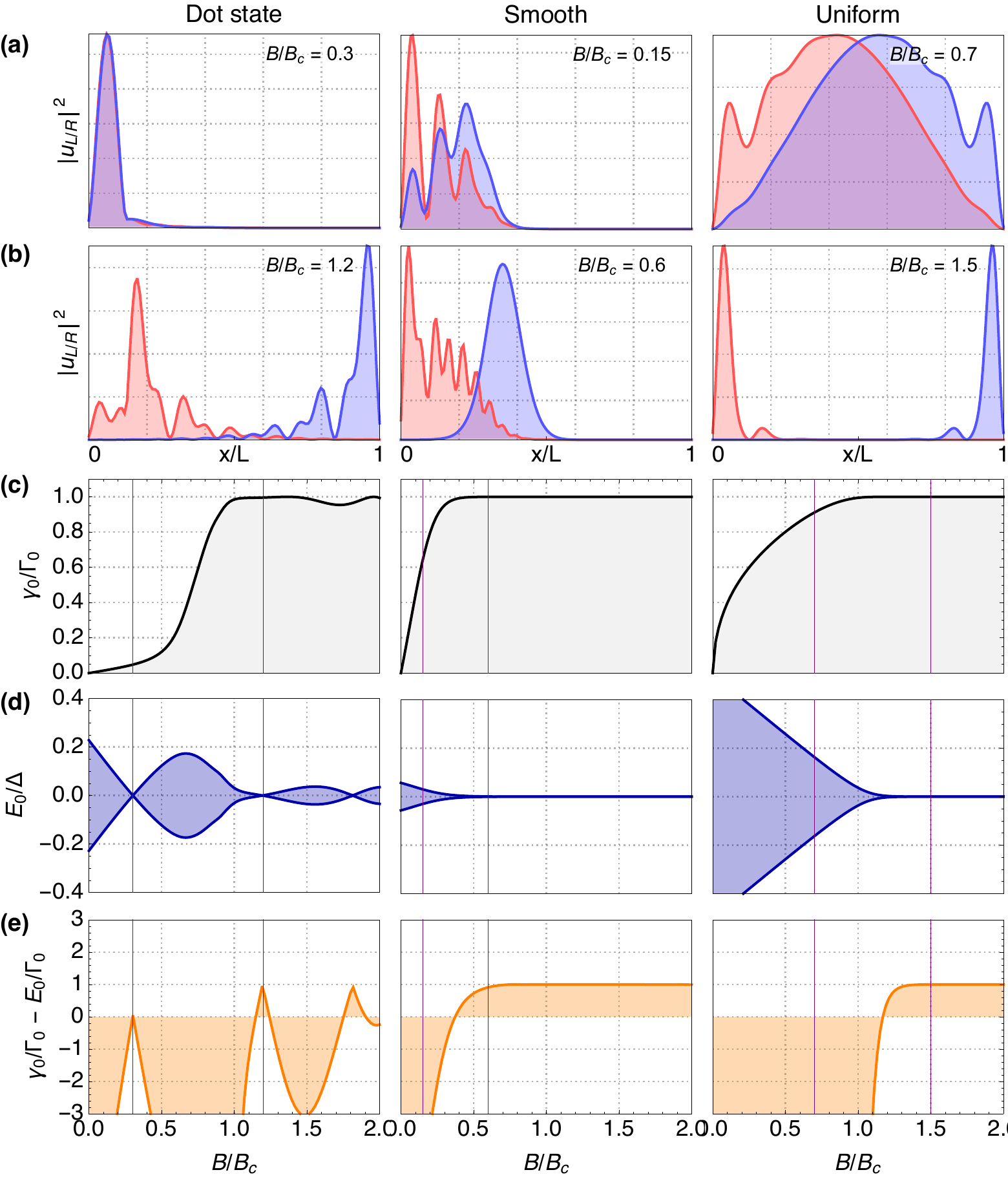}
\caption{\label{Fig2 SI} \textbf{Majorana non-locality and coupling asymmetry}. (a) and (b)  Majorana wave functions along the nanowire for the three systems of Fig. \ref{Fig2}: short wire with a quantum dot state in the N region, nanowire with a smoothly confined ABS and a nanowire with uniform density and pairing. (a) shows low Zeeman field $B$ state with strongly overlapping Majorana wave functions, while (b) show how non-local Majorana components develop at higher $B$. The corresponding $B$ are marked with vertical lines in panels (c-e), where we show the $B$ dependence of the coupling asymmetry, the energy $E_0$ and the criterion for EP formation. Sign changes in the normalised difference $(\gamma_0-|E_0|)/\Gamma_0$  (panels in e) appear when \emph{both} the energy is close to zero (panels in d) and there is sufficient wave function non-locality $\gamma_0/\Gamma_0$ (panels in c). These sign changes at discrete Zeeman fields mark the appearance of EPs and quasi-bound Majorana zero modes, and thus constitute a topological criterion por open NS nanowire junctions.}
\end{figure}
\begin{figure} 
\centering \includegraphics[width=\columnwidth]{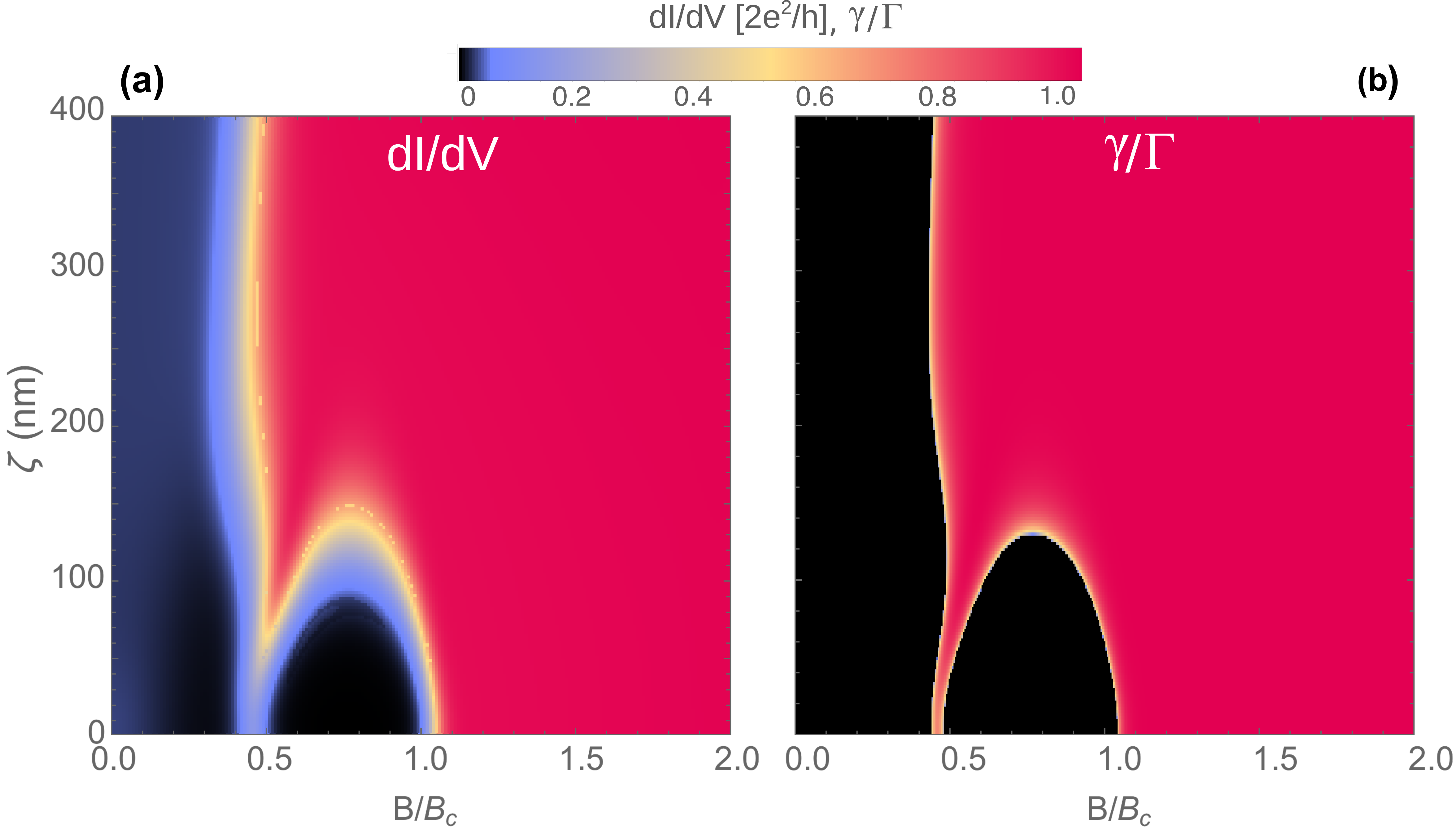}
\caption{\label{Fig 3SI} \textbf{Phase diagram of smooth NS junction}. (a) Linear conductance $dI/dV|_{V\to 0}$ at T=20mK as a function of Zeeman field and smoothness parameter $\zeta=\zeta_N=\zeta_S$. As smoothness increases, the conductance quickly saturates to $2e^2/h$ already for $B\gtrsim 0.5B_c$. (b) The jump in the decay asymmetry from a trivial value $\gamma/\Gamma\approx 0 $ to a non-trivial one $\gamma/\Gamma\approx 1$ follows the same pattern. Both phase diagrams demonstrate the generality of the EP phenomenon in smooth NS junctions to generate non-trivial Majorana zero modes for $B<B_c$.}
\end{figure}
\subsection{Wave function non-locality and coupling asymmetry as the origin of exceptional points in NS junctions} As we argue in the main text, so called trivial Andreev levels may bifurcate into quasi-bound zero modes, exactly like conventional Majorana zero modes, as long as the wavefunctions of the zero-mode's Majorana components become spatially separated in such a way that the 
coupling asymmetry to the reservoir  is larger than their energy $\gamma_0>|E_0|$. We demonstrate this claim in Fig. \ref{Fig2 SI}, where we plot wave functions of the Majorana components for the three geometries discussed in the main text. The panels in Fig. \ref{Fig2 SI}a show three examples with strongly overlapping Majorana wave functions, while the three panels Fig. \ref{Fig2 SI}b show examples of non-local Majorana components. The corresponding Zeeman fields are marked with vertical lines in panels (c-e). The smooth NS geometry (central panel in Fig. \ref{Fig2 SI}b) produces spatially separated Majorana components for moderate Zeeman fields ($B=0.6B_c$ in the example), which correspond to a very robust zero energy state (central panel in Fig. \ref{Fig2 SI}d). As we argue in the main text, this Majorana non-locality implies that the coupling asymmetry becomes maximal, $\gamma_0/\Gamma_0=1$, \emph{well before} the bulk topological transition $B=B_c$. We demonstrate this effect in 
Fig. \ref{Fig2 SI}c where we plot the asymmetry ratio 
\begin{eqnarray}
\label{gamma}
\frac{\gamma_0}{\Gamma_0}&\equiv&\frac{\langle u_L|H_\mathrm{eff}|u_L\rangle-\langle u_R|H_\mathrm{eff}|u_R\rangle}{\langle u_L|H_\mathrm{eff}|u_L\rangle+\langle u_R|H_\mathrm{eff}|u_R\rangle}\nonumber\\
&=&\frac{\langle u_L|\Sigma_{x=0}|u_L\rangle-\langle u_R|\Sigma_{x=0}|u_R\rangle}{\langle u_L|\Sigma_{x=0}|u_L\rangle+\langle u_R|\Sigma_{x=0}|u_R\rangle}\nonumber\\
&=&\frac{|u_L(0)|^2-|u_R(0)|^2}{|u_L(0)|^2+|u_R(0)|^2},
\end{eqnarray}
as a function of Zeeman field. Here, $H_\mathrm{eff} = H_0+\Sigma_{x=0}$ and $\Sigma_{x=0}$ is the self-energy introduced on the first site at $x=0$ by the coupling to the reservoir, while $u_{L,R}(x) = \langle x|u_\mathrm{L,R}\rangle$ are the wavefunctions of the Majorana components $|u_\mathrm{L,R}\rangle$ of the two lowest eigenstates of $H_0$, the discretized Oreg-Lutchyn Hamiltonian of the decoupled system.
The smooth case (central panel) indeed shows how the ratio $\gamma_0/\Gamma_0$ smoothly crosses over from 0 to 1 at Zeeman fields $B<B_c$.
Conversely, the other cases (Fig. \ref{Fig2 SI}c, left and right panels) only develop full coupling asymmetry $\gamma_0/\Gamma_0=1$ after a standard topological transition at $B=B_c$. Said asymmetry leads to the development of an EP bifurcation as soon as the criterion $\gamma_0>|E_0|$ is fulfilled (irrespective of whether a bulk band-topological transition has ocurred). This is demonstrated in panels Fig. \ref{Fig2 SI}e, where sign of $(\gamma_0-|E_0|)/\Gamma_0$ changes \emph{exactly} at values of the Zeeman field where the real energy of the lowest poles in the coupled system is stabilised to zero, as a result of sufficient Majorana non-locality. Note that, despite the fact that the non-locality ratio $\gamma_0/\Gamma_0$ changes smoothly with Zeeman, the criterion $(\gamma_0-|E_0|)/\Gamma_0$ changes sign at \emph{discrete} Zeeman fields which mark the topological EP biburcation point where nontrivial Majorana zero modes appear. We emphasize that a trivial parity crossing with zero energy but $\gamma_0/\Gamma_0\approx 0$ does \emph{not} result in a sign change (left panel in Fig. \ref{Fig2 SI}e at $B=0.3B_c$) and thus does not bifurcate through an EP (see Fig. \ref{Fig2}d). We want to stress that, although they are in perfect agreement with the physics provided by the low energy model in the main text, these results have been obtained \emph{without} any use of Eq. (\ref{EP}), just from the couplings in Eq. (\ref{gamma}) which involve full microscopics of the studied models. This demonstrates the main claim of this paper and constitutes a proof that, indeed, EPs linked to wave function non-locality are the correct topological criterion in NS junctions, instead of that governed by bulk band topology.

\subsection{Topological phase diagrams using the EP criterion}

As we discuss in the main text, ABSs confined in smooth NS junctions become stabilised at zero real energy through an EP bifurcation of its two Majorana decay rates. This gives rise to nontrivial Majorana physics \emph{well before} the critical Zeeman field $B_c$. To investigate more systematically these EP transitions due to smooth NS interfaces, \editP{we study here the linear conductance $dI/dV|_{V\to 0}$, computed at T=20mK, as a function of Zeeman field $B$ and contact smoothness $\zeta$, see Fig. \ref{Fig 3SI}a. It shows two distinct regions: a trivial one with low conductance (black), and a non-trivial one with nearly-quantized $2e^2/h$ (red). The latter begins at $B\approx 0.5B_c$ for smoothness $\zeta$ exceeding a threshold of the order of the Fermi wavelength. As in Fig. 2 of the main text, we compare the onset of quantised conductance to the corresponding decay asymmetry $\gamma/\Gamma$ (right panel) in the same parameter space. As in Fig. 2 we clearly see the correlation between the two quantities, which validates our interpretation of $\gamma/\Gamma$ as the relevant non-Hermitian topological order parameter associated to non-trivial phenomenology.}
In all cases studied, point-like parity crossings in systems with sharp interfaces  $\zeta\rightarrow 0$ (e. g. at $B\approx 0.5B_c$ in the density plots) quickly evolve into stable Majorana zero modes as $\zeta$ increases. For sufficient $\zeta$ they appear in the corresponding $B-\zeta$ phase diagrams as extended areas with quantised conductance $2e^2/h$ and a maximum asymmetry $\gamma/\Gamma\approx 1$.

\bibliography{biblio}

\end{document}